\documentclass[twocolumn]{aastex631}
\usepackage{xcolor}

\newcommand{\kb}{k_{\text{B}}}

\usepackage{color, colortbl}
\definecolor{Gray}{gray}{0.75}
\usepackage{graphicx}
\usepackage{hyperref}
\usepackage{natbib}
\usepackage{amssymb}
\usepackage{amsmath}
\usepackage{relsize}
\usepackage{url}
\usepackage{color, colortbl}
\usepackage{enumerate}
\usepackage{comment}
\usepackage{ulem}
\usepackage[version=4]{mhchem}

\newcommand\avtz{\textsf{avtz}}
\newcommand\avqz{\textsf{avqz}}
\newcommand\avcbs{\textsf{avcbs}}
\newcommand\avcz{\textsf{av5z}}
\newcommand\COTwo{{${\mathrm{CO_2}}$}}

\begin{document}

\title{\textsl{Ab initio} quantum dynamics as a scalable solution to the exoplanet opacity challenge:
A case study of CO$_2$ in hydrogen atmosphere}

\author[0000-0003-2355-4543]{Laurent Wiesenfeld}
\affiliation{Universit\'e Paris-Saclay, CNRS, Laboratoire Aim\'e-Cotton, 91405 Orsay, France }
\affiliation{Department of Earth, Atmospheric and Planetary Sciences, MIT, 77 Massachusetts Avenue, Cambridge, MA 02139, USA}
\email{Email: laurent.wiesenfeld@universite-paris-saclay.fr}

\author[0000-0002-8052-3893]{Prajwal Niraula}
\affiliation{Department of Earth, Atmospheric and Planetary Sciences, MIT, 77 Massachusetts Avenue, Cambridge, MA 02139, USA}
\email{Email: pniraula@mit.edu}

\author[0000-0003-2415-2191]{Julien de Wit}
\affiliation{Department of Earth, Atmospheric and Planetary Sciences, MIT, 77 Massachusetts Avenue, Cambridge, MA 02139, USA}

\author{Nejmeddine Jaïdane}
\affiliation{Université Tunis El Manar, Faculty of Sciences, Tunis, Tunisia}
\affiliation{Universit\'e Paris-Saclay, CNRS, Laboratoire Aim\'e-Cotton, 91405 Orsay, France }

\author[0000-0003-4763-2841]{Iouli E. Gordon}
\affil{Harvard-Smithsonian Center for Astrophysics, Atomic and Molecular Physics Division, Cambridge, MA, USA}

\author[0000-0002-7691-6926]{Robert J. Hargreaves}
\affil{Harvard-Smithsonian Center for Astrophysics, Atomic and Molecular Physics Division, Cambridge, MA, USA}


\begin{abstract}
Light-matter interactions lie at the heart of our exploration of exoplanetary atmospheres. Interpreting data obtained by remote sensing is enabled by meticulous, time- and resource-consuming work aiming at deepening our understanding of such interactions (i.e., opacity models). Recently, \citet{Niraula2022} pointed out that due primarily to limitations on our modeling of broadening and far-wing behaviors, opacity models needed a timely update for exoplanet exploration in the JWST era, and thus argued for a scalable approach. In this proof-of-concept study, we introduce an end-to-end solution from \textsl{ab initio} calculations to pressure broadening, and use a perturbation framework to identify the need for precision to a level of $\sim$10\%. We focus on the CO$_2$-H$_2$ system as CO$_2$ is a key absorption feature for exoplanet research (primarily in many gas giants) at $\sim$4.3$\mu$m as pressure-broadening parameters required for interpreting such observations remain sparse. We compute elastic and inelastic cross-sections for the collision of {ortho-}H$_2$~with CO$_2$, in the ground vibrational state, and at the coupled-channel fully converged level. For scattering energies above $\sim$20~cm$^{-1}$, moderate precision inter-molecular potentials are indistinguishable from high precision ones in cross-sections. Our calculations agree with the currently available measurement within 7\%, i.e., well beyond the precision requirements.
\end{abstract}

\color{black}

\section{Introduction}
\label{sec:intro}
Amongst the JWST molecular observations, carbon dioxide is almost universally observed in exoplanetary atmospheres. From hot Jupiters such as WASP-39 b \citep{rustamkulov2023, Niraula2023} to temperate sub-Neptunes such as K2-18~b \citep{madhusudhan2023}, transmission spectra have shown the characteristic spectral feature at 4.3~microns. In fact, as this feature falls in a spectral sweet spot where the overall noise budget and the effect of cloud and/or hazes are minimal, it is also ideal for the detection of temperate terrestrial atmospheres and assessing their (in)habitability \citep[e.g.,][]{triaud2023,TJCI2024}.

\begin{figure*}[ht!]
    \centering
    \includegraphics[trim={0cm 2cm 0cm 2cm},clip,width=0.95\textwidth]{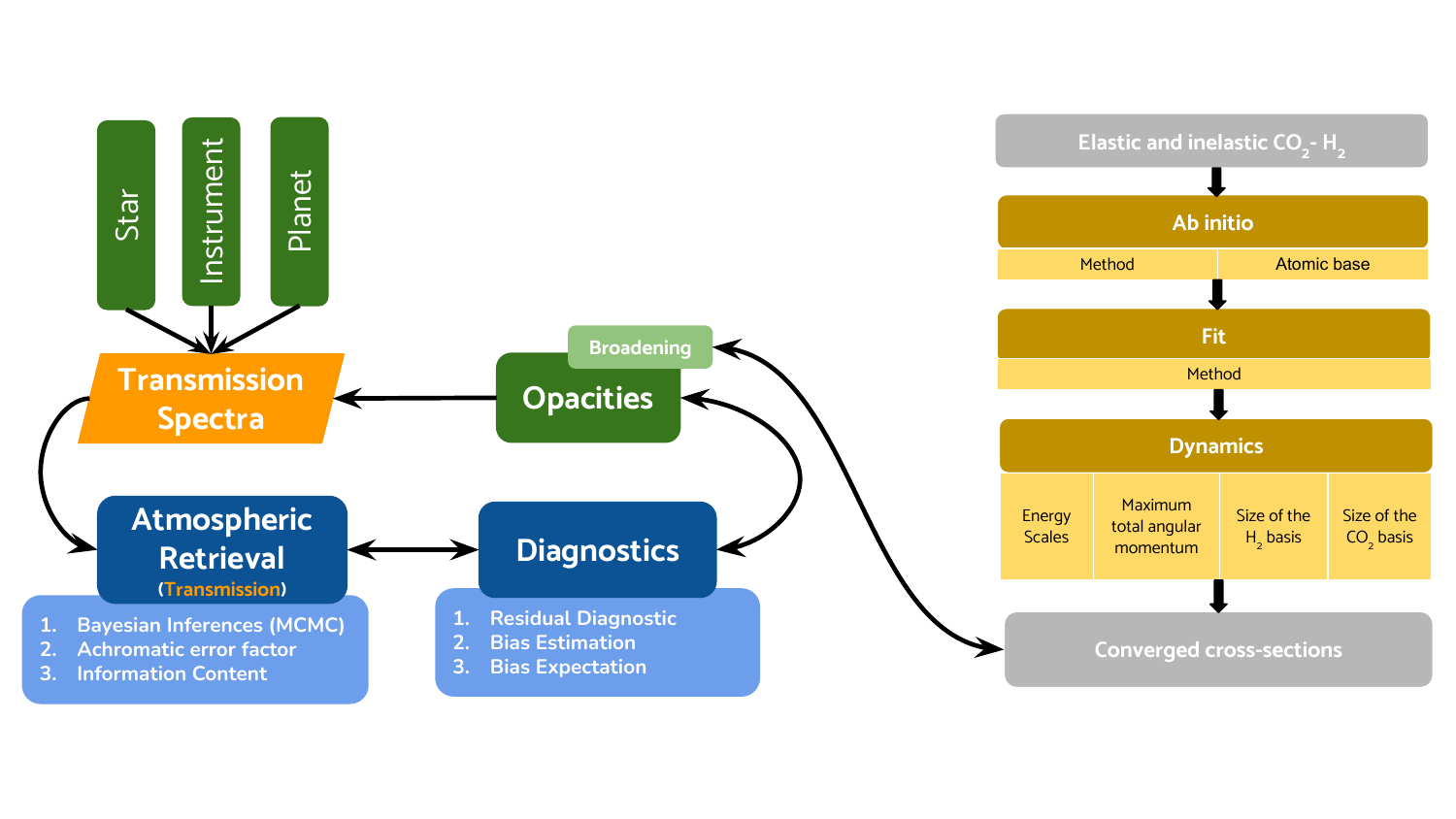}
    \caption{Flowchart for exoplanet atmosphere retrieval focusing on broadening parameters, as a key source of limitations in current opacity models (adapted from N22). \textsl{Ab initio} calculations are performed with the \textsc{MOLPRO} code \citep{MOLPRO_brief},  fits: Rist (\cite{Rist:2012aa} and \textsc{YUMI}  for scattering and dynamics code (Jaïdane et al. in prep.). }
    \label{fig:scheme}
\end{figure*}

Yet, despite these numerous favorable aspects, exploring exoplanet atmospheres using CO$_2$ can be challenging due to the current state of its opacity model. \citet{Niraula2022} (henceforth, N22) pointed out that the current limitations of opacity models (primarily due to broadening and far-wing parameterizations) can yield an accuracy wall on the inferred atmospheric properties at the level of $\sim$1\,dex, orders of magnitude above JWST's capabilities (see \autoref{fig:scheme} for general framework). \cite{Niraula2023} subsequently highlighted that the aforementioned 4.3~$\mu$m feature of CO$_2$, while dominant in most exoplanet spectra, can be challenging to translate into robust atmospheric inferences for a large range of exoplanets as these will harbor a hydrogen-dominated atmosphere, and very scarce data exists for broadening of CO$_2$ lines by collisions with H$_2$, especially at the higher temperatures \citep{hanson2014,padmanabhan2014}. This is partly because the experiments on carbon dioxide and molecular hydrogen mixtures, especially at elevated temperatures, are very challenging from a safety perspective, adding extra complexity and expense to already existing difficulties in accurate measurements of broadening parameters.  Due to a lack of data, the HITRAN spectroscopic database \cite{hitran2020,2022ApJS..262...40T}, for instance, uses a very approximate solution of scaling the well-known air-broadening data based on limited hydrogen-broadening data. Since the dependence of air and hydrogen broadening data on the type of the transition is very different, the errors associated with this approach can be quite large.

In the light of the hurdles in providing a solution to the opacity challenge on a timescale relevant to the JWST mission, this study presents the viability of a new-generation framework for \textsl{ab initio} calculations with a focus on pressure broadening coefficients (typically defined as half-widths at half maximum (HWHM) \cite{hitran2020}). Different theoretical methods for calculating the pressure broadening employing classical dynamics/semi-classical quantizing exist, with various degrees of approximation \citep{2018JQSRT.213..178H,Ngo:2021aa}.  For example, \citet{2024Buldyreva} have used classical phase-shift theory to provide some data/estimates for the UV/Visible range. There is also a large series of classical computations using very approximate potentials, but capable of computing the whole lineshape, including the far wings \citep{HARTMANN2002117}. Also, \citet{2024Guest} have investigated machine learning to provide broadening data based on HITRAN parameters. This is one avenue that would be considered state-of-the-art, but is limited by the learning set and will struggle to predict broadening for molecules/systems not included.
In other words, while the \textsl{ab initio} calculation is not the only theoretical approach to address the opacity driven accuracy wall (a wall which is expected to be hit once the retrieved precision surpasses 1 dex (see Fig. 5 in \citet{Niraula2022}), it is certainly one of the most accurate (see, for instance, \cite{2021JQSRT.27207807S, Gancewski2021}). And, as shown here, it can provide a cost-effective and thus scalable solution that can be deployed in a timescale relevant for ongoing missions such as JWST.

First, we assess the precision requirement on the pressure broadening calculations given JWST's capabilities in \autoref{sec:requirements}. Then, in \autoref{sec:methods}, we describe our method for calculating the pressure broadening parameters for CO$_{2}$+H$_{2}$. Our results are presented in \autoref{sec:results}. We discuss implications (incl., scalability) and conclude in \autoref{sec:conclusion}. The necessary background pertaining to the pressure broadening calculations is included in the \autoref{sec:quantum}.

\section{Precision Requirements}
\label{sec:requirements}

In this section, we aim to provide a more granular understanding of the pressure-broadening contribution to the uncertainty budget in exoplanet atmosphere analysis than N22 in order to set the precision requirement on the pressure broadening calculations. Following the approach of \citet{Berardo2024} and \citet{Rackham2024}, we require that the opacity perturbations contribute to the uncertainty budget on the atmospheric properties derived no more than the measurement itself, ensuring instrument-limited science.


To this end, we perform cross-retrievals with perturbed cross-sections on the two synthetic planetary systems following the framework introduced in N22---one super-Earth (1.1 R$_\oplus$, 1.1 M$_\oplus$ around a 0.1R$_{\odot}$ M-dwarf star) and a warm Jupiter (1.0 R$_{\rm Jup}$, 1.0 M$_{\rm Jup}$ around a 0.55R$_{\odot}$ K-dwarf star). We use \texttt{pandexo} \citep{pandexo} to model instrumental precision using the same setup as in N22.  As we focus here on the influence of perturbations on pressure broadening coefficients to test the sensitivity of atmospheric inferences on that single parameter, we use a simpler synthetic atmosphere than in N22, considering only four molecular species (hydrogen, helium, carbon dioxide, and water). For the super-Earth scenario, we consider a high metallicity (MR H$\mathrm{_2}$O=0.1, MR CO$\mathrm{_2}$=0.1) isothermal (T=300 K) cloudless atmosphere, whereas for the warm-Jupiter scenario we consider a slightly hotter low-metallicity atmosphere (T=400 K; MR H$\mathrm{_2}$O=$2\times10^{-4}$, MR CO$\mathrm{_2}$=$1\times10^{-4}$).

\begin{figure*}[ht!]
    \centering
     \includegraphics[trim={1.5cm 0.5cm 0cm 1cm},clip, width=0.95\textwidth]{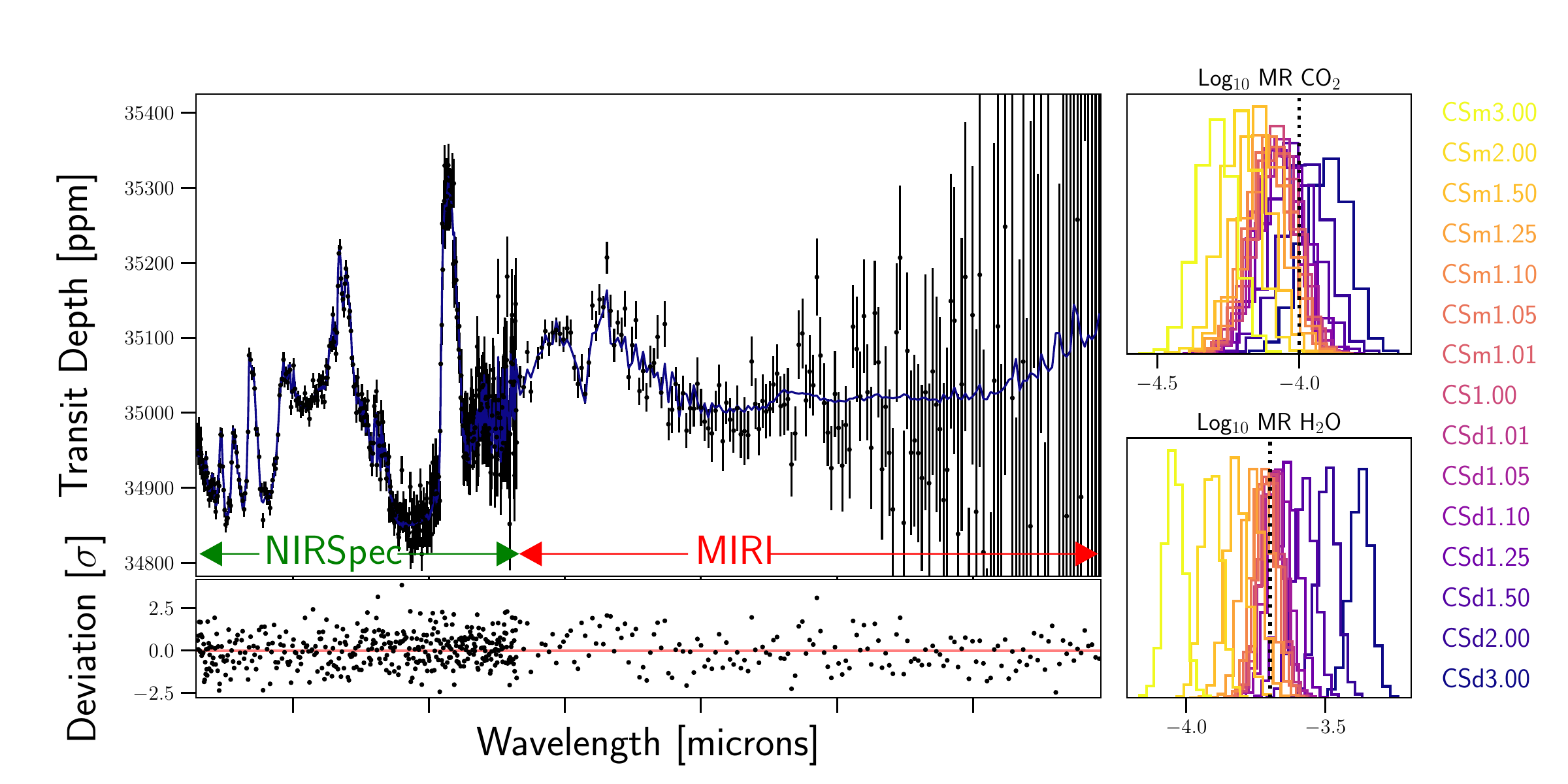}
    \caption{Perturbation analysis via injection-retrieval to assess the precision required on the broadening parameters in the JWST era. \textbf{Top Left:} Synthetic transmission spectrum of a warm-Jupiter spanning NIRSpec and MIRI bands used for the retrieval. The blue line represents the best-fit model obtained when cross-retrieved with the CSd3.00 cross-sections. \textbf{Bottom Left:} Deviation of the residual corresponding to the best fit presented on the upper panel. No structure is seen due to parameter compensation as highlighted in N22. \textbf{Top Right:} Posterior probability distributions for the mixing ratio of CO$_{2}$ for 15 different cross-sections showing increased levels of biases with increased perturbation. The dotted black line represents the `truth' used for generating the synthetic spectrum. Overall variation of $\sim$1 dex is seen among the retrieved molecular abundance, consistent with N22. \textbf{Bottom Right:} Same for $\mathrm{H_2O}$. A similar variation of $\sim$1 dex overall, though leading to more significant biases due to the narrower distributions than for $\mathrm{CO_2}$.  }
    \label{fig:retrieval}
\end{figure*}

To estimate the perturbation level leading to biases of amplitude equal to the 1-$\sigma$ confidence interval on atmospheric inference given by the instrumental precision, we generate an ensemble of perturbed cross-sections (CS, hereafter). Perturbations relate here to the underlying pressure-broadening parameters, and the series of cross-sections used are denoted by CSxYYY where x is either m (for multiplied) or d (for divided) and 1.00$\leq$YYY$\leq$3.00, thereby referring to the scaling of the pressure-broadening parameters. This ensemble includes fifteen relevant cross-sections. The nominal cross-section, CS1.00 was produced using the best available parameters from HITRAN database, generated by prescription that included scaling of the air-broadening parameters. The rest fourteen cross-section were perturbed accordingly: CSd3.00 (i.e., all the broadening parameters are one-third of those reported as nominal), CSd2.00, CSd1.50, CSd1.25, CSd1.10, CSd1.05, CSd1.01, CSm1.01, CSm1.05, CSm1.10, CSm1.25, CSm1.50, CSm2.00, and CSm3.00 (i.e., all the broadening parameters are three times those reported as nominal).  As in N22, we use \texttt{HITRAN} \citep{hitran2020}/\texttt{HITEMP} \citep{HITEMP2010} for our opacity model inputs.
We then performed the injection-retrieval tests, generating the synthetic spectra using the unperturbed cross-section (CS1) and performing the retrieval with all cross-sections. We ran our atmospheric models with \texttt{emcee} for a step-size of 5000, ensuring our walkers are fully converged.

\autoref{fig:retrieval} shows the typical results derived here for the warm Jupiter scenario. As found in N22, even for large perturbations on the cross-sections, a good fit to the data can be found (no structure in the residuals) owing to compensation by the atmospheric parameters in the associated dimensions of the inverse problem. It serves as a reminder that obtaining a good fit does not  guarantee confidence in our inferences (and models). In fact, it further stresses the need for efforts such as this study to produce models of sufficient, independently-tested fidelity to guarantee that the inferences will be reliable. Indeed, without such independent test of the model fidelity nor the capability to leverage the quality of the fit (or lack thereof) to assess it, one could reach inferences suffering unknowingly---a key takeaway from N22. These compensations translate into biased inferences up to $\sim$1~dex, for example, on the abundance of absorbers such as carbon dioxide and water (right panels of \autoref{fig:retrieval}). While both water and carbon dioxide showcase biases up to $\sim$1~dex, the statistical significance of the biases for water appears more prominent due to the narrower posterior probability distribution, which results from the higher information content associated with water (primarily due to the wider wavelength range associated with its absorption features).

\begin{figure}
    \centering
     \includegraphics[trim={0cm 0.75cm 0cm 0.0cm},clip, width=0.475\textwidth]{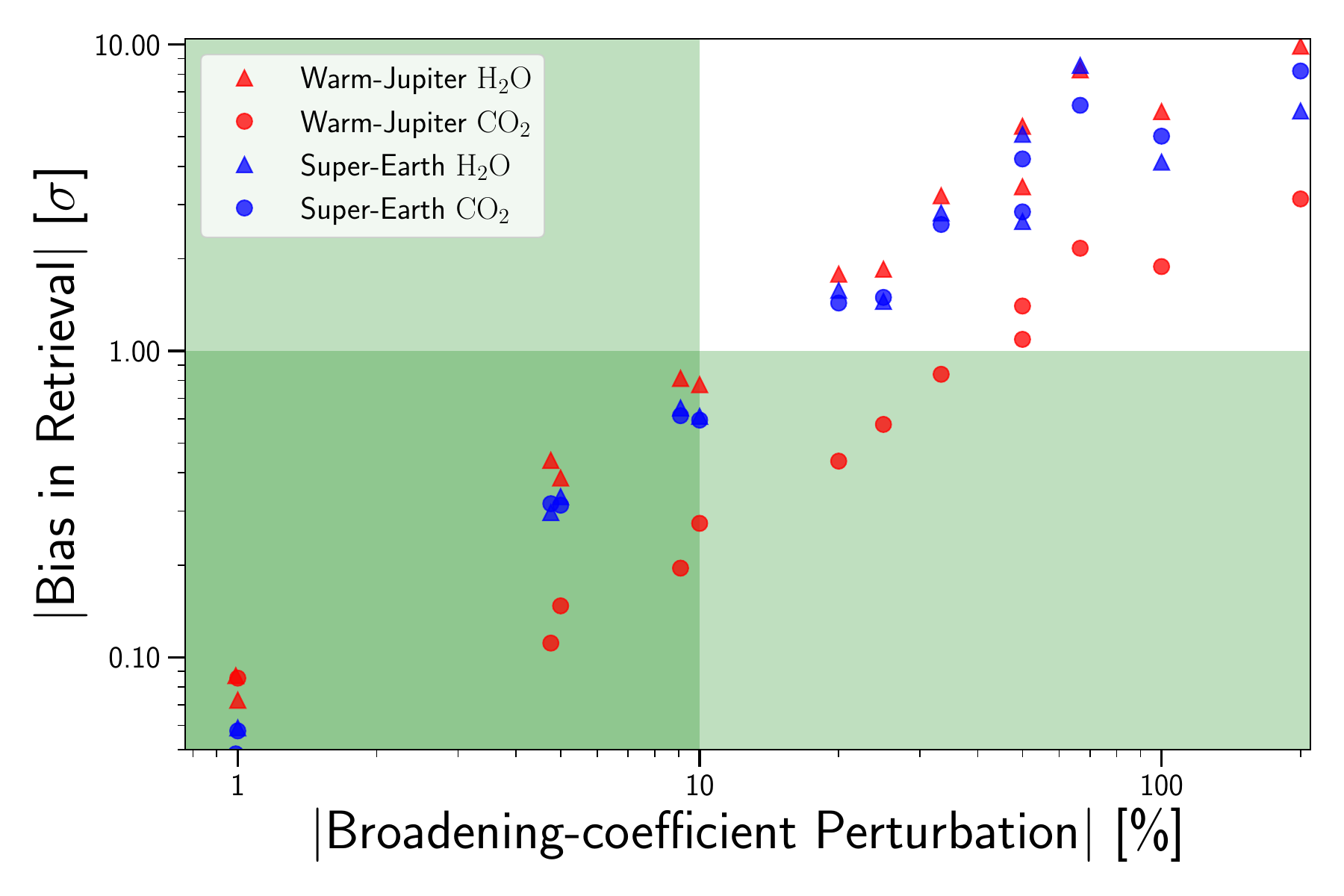}
    \caption{ Greater than 5$\sigma$ biases observed among the retrieved water and carbon-dioxide abundances in our retrievals, with larger perturbations inducing stronger biases. The green region represents an acceptable bias range from -1$\sigma$ to +1$\sigma$, demonstrating a need for precision of 10-15\% in the pressure broadening parameter for JWST observations.}
    \label{fig:perturbationBias}
\end{figure}

We summarize our precision-requirement tests in \autoref{fig:perturbationBias}. As expected (and as shown in the right panels of \autoref{fig:retrieval}), it shows that the biases on the atmospheric inferences decrease with the perturbation level. The biases in \autoref{fig:perturbationBias} are reported in significance level ($\sigma$), i.e., the ratio between the bias and the width of the distribution which is solely driven by the measurement uncertainty. It shows that biases become marginal (i.e., below 1$\sigma$) when the perturbation on the broadening parameters drops below $\sim$10\%. In other words, our precision requirement for broadening parameters to ensure instrument-limited science in the JWST era (i.e., avoid model-limited inferences) is 10\%  (see \citet{Niraula2022} for bias mechanisms, and see \citet{Niraula2023} for the application to WASP-39 b).

\section{Methods}
\label{sec:methods}

With the well-justified precision requirement in hand, we can now turn to the series of calculations required to derive \textsl{ab initio} estimates of the broadening parameters. In this section, we provide an overview of the steps. For further details, refer to the  \autoref{sec:quantum}.

\subsection{Potential Energy Surface}

We begin by computing the electronic interaction energy of the two molecules taken both as rigid rotors, employing the \textsl{ab initio} \textsc{MOLPRO} code \citep{MOLPRO}. We solve the electronic Schr\"odinger equation with fixed nuclei, using a quantum chemistry method CCSD(T) that allows for a good description of the electronic correlations, even for two molecules separated by a sizeable distance (here, up to 25 \AA). The potential energy surface (PES) for the \ce{CO2-H2} interaction was computed for some 500 relative orientations of the molecules and for 29 intermolecular distances ranging from approximately 1.75~{\AA} to 25~{\AA}. For shorter distances, the two molecules repel so strongly ($E_{\text{interaction}} \gtrsim 10,000 \, \mathrm{cm}^{-1}$) that the approximation of two isolated molecules does not hold. For distances larger than about 25~\AA, the interaction is weak ($E_{\text{interaction}}\lesssim 1 \, \mathrm{cm}^{-1}$) and becomes neither reliable nor relevant for our computations.  Note that because of the symmetry of both molecules ($D_{\infty h}$), the potential is $\pi$-periodic in the dihedral angle $\phi$ and $\pi/2$-periodic in both polar angles $\theta_1$ and $\theta_2$ (see \autoref{fig:geometry}).

\begin{figure*}[!ht]
    \centering
    \includegraphics[trim={0cm 9cm 0cm 6cm},width=0.85\textwidth]{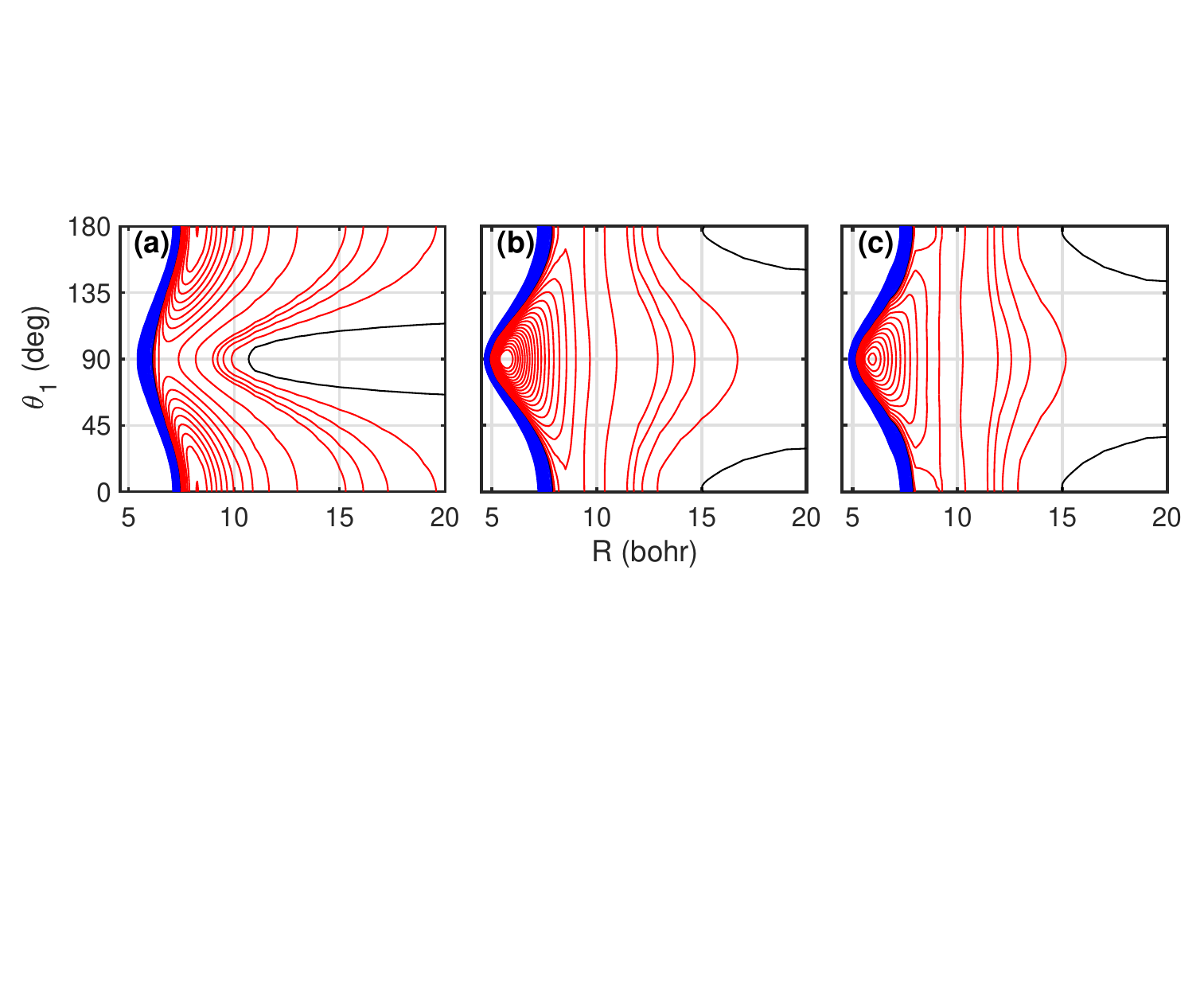}
    \caption{Three 2-dimensional cuts of the 4-dimensional interaction potential \ce{CO2-H2}. Planar configurations: panel (a) : $\theta_2=0,\,\phi=0$; panel (b) $\theta_2=\pi/2,\,\phi=0$; non-planar configuration: panel (c) $\theta_2=\pi/2,\,\phi=\pi/2$. Blue lines represent positive values of the potential, red lines negative values, and the black lines zero values.  Note that the potential is only asymptotically isotropic. }
    \label{fig:2D_cuts}
\end{figure*}

We then fit the \textsl{ab initio} isolated points onto a functional form according to the geometrical conventions of the angles $\phi, \theta_1, \theta_2$, see equation (A9) of \citep{GREEN:1975ae} and \autoref{fig:geometry}. We performed several fits of various precision, with 38 up to 158 terms, and found that an acceptable precision ($|V_{\text{fit}}-V_{\text{\textsl{ab initio}}}|/|V_{\text{fit}}+V_{\text{\textsl{ab initio}}}|\leq $ a few \%) is reached as soon as the orientation of the H$_2$ molecules is described with at least $l_2=0,2,4$ $Y_{l_2,m_2}(\theta_2,\phi$) spherical harmonics. Ultimately, we used a 149 terms expansion, up to $l_1=24, \, l_2=6$ terms in $Y_{l_1,m_1}(\theta_1,0)Y_{l_2,m_2}(\theta_2,\phi$). 2-D cuts of the potential energy surface are presented in \autoref{fig:2D_cuts}, emphasizing the change of the shape of the PES as the orientation of \ce{H2} changes.

\subsection{Dynamics: Broadening Coefficient}
With the PES fits in hand, we can solve the quantum scattering dynamics in the time-independent domain, using the close-coupling formalism \citep{GREEN:1975ae, valiron2008}.  We use a home-written code  (YUMI), specifically designed for taking advantage of massively parallel computing. The code has been extensively tested against the existing state-of-the-art MOLSCAT code \citep{Hutson2019}, and will be described in detail in a forthcoming paper (Jaïdane et al., in prep.).

The quantum computations of the pressure-broadening cross-section $\sigma_{\text{PB}}(E)$ and the rate coefficient $\gamma(T)$ are performed in the approximation of isolated binary collisions. It allows for no overlap between the different spectral lines: These hypotheses remain valid up to about one amagat of density (about $10^{19}$  particles/cm$^3$) and for P or R branches of the IR spectrum, but not for the Q branches. The isolated collision hypothesis allows for computing first the $\mathsf{S}(E)$ matrices of the collision and derive from these the desired cross-sections, whether elastic, inelastic or pressure broadening and pressure shift \citep{GREEN1977119,Drouin:2012aa,Faure:2013aa,Selim:2023aa}. The cross-sections are computed at given collision energies (the relative kinetic energy of the two molecules), then averaged over a Maxwellian distribution of kinetic energies to yield the relevant rates.

For this proof-of-concept application, we focus on reproducing the \textsl{few experimentally} known values, measured at different temperatures \citep{hanson2014,padmanabhan2014},  which connects an asymmetric stretch excited state with the ground state. The \citet{hanson2014} measurement targets P(24) transition of the 20012-00001 band, whereas \citet{padmanabhan2014} covers P(16)-P(34) transitions of the 30012-00001 band. (For spectroscopic notations, see section \ref{sec:notations}).
Based on selection rules,  the only allowed transitions are $v'_3=1,\,j_1' \text{~odd} \leftrightarrow v''_3=0,\ j_1''=j_1'\pm 1$. The nuclear spin statistics dictates that only even rotational levels in the ground vibrational state are populated because $\mathrm{^{12}C^{16}O_2}$ is a symmetric molecule with zero nuclear spin values for all atoms, hence exist only for even  $j_1$ in the ground state and odd $j_1$ in the $v_3=1$ case. The $v_1$ excitation does not interfere with symmetry rules, being fully symmetric, merely acting as a spectator to the IR transition as demonstrated in the experimental works \citep{hanson2014, padmanabhan2014} (Spectroscopic notations, see Sec. \ref{sec:notations}).

The experimental values for the P24 ($m=-24$), in cm$^{-1}$/atm, are $\gamma = 0.112$, for \citep{hanson2014} and $\gamma = 0.113 \pm 0.006$, for \citep{padmanabhan2014}, emphasizing the weak dependence on the exact vibrational transition. Errors bars for \cite{hanson2014} are of the order of a few \%.

It has recently been shown \citep{Wiesenfeld:2022aa,2021JPCA..125.6864Y} that computing inelastic scattering cross-sections between two different ro-vibrational states is doable, but requires extremely large computational resources, which would render the scalability of our approach questionable. Fortunately, the pressure broadening effect arises through \emph{quasi-elastic} collision amplitudes,
connecting the $j'_2,v'_3$ or $j''_2, v''_3$ levels with themselves, but for different orbital angular momenta, see section \ref{sec:pbcalc}. The experimental consequence is a very weak dependence of the pressure broadening on the molecular vibrations, except for some very anharmonic vibrations of modes containing atom(s) of hydrogen. The theoretical consequence is that we are allowed to consider the interaction potential and the dynamics as depending only on the average geometry of the molecules --rigid rotors approximation--, and not on the vibrational dynamics, making the whole computation scalable. If selection rules allow (e.g., molecules of $C_{\infty h}$ symmetry with no plane of symmetry perpendicular to the axis like CO or HCN ), the pure rotational transitions serve as a proxy for the actual IR transitions (see e.g. \cite{Dumouchel:2010aa,1989JChPh..90.3603G,Luo:2001aa}). For the case of molecules with  $D_{\infty h}$  (e.g., with a plane of symmetry perpendicular to the axis like \COTwo~or C$_{2}$H$_{2}$), the rotational selection rules differ from the vibrational selection rules. For \COTwo, only even rotational levels exist for the vibrational ground state and only odd rotational levels exist for the asymmetric stretch first excited levels ($v_3$ = 1). We used these levels to compute pressure broadening for the P and R branches in the IR transition. Calculations by \cite{Rosenmann1988} show that there is practically no difference between the broadening of IR and Raman transitions of CO$_2$ for several different projectile molecules, and they also did not observe the vibrational dependence. (Note: “Target” and “projectile” molecules are standard denominations in scattering studies are respectively refer to the observed molecule---the target, and the bath molecule---the projectile, or broadener.)  In general, in examining experimental data for linear molecules in the  HITRAN database \citep{hitran2020,2022ApJS..262...40T,hashemi2020}, weak or no dependence on vibrational modes is typically observed for linear molecules.

\section{Results}
\label{sec:results}

As previously mentioned, our goal via this proof-of-concept of a scalable \textit{ab initio} framework for broadening-parameter generation is to reproduce the \textsl{only experimentally} known values \citep{hanson2014,padmanabhan2014}, measured at different temperatures.

\begin{figure}
    \centering
    \includegraphics[width=0.475\textwidth]{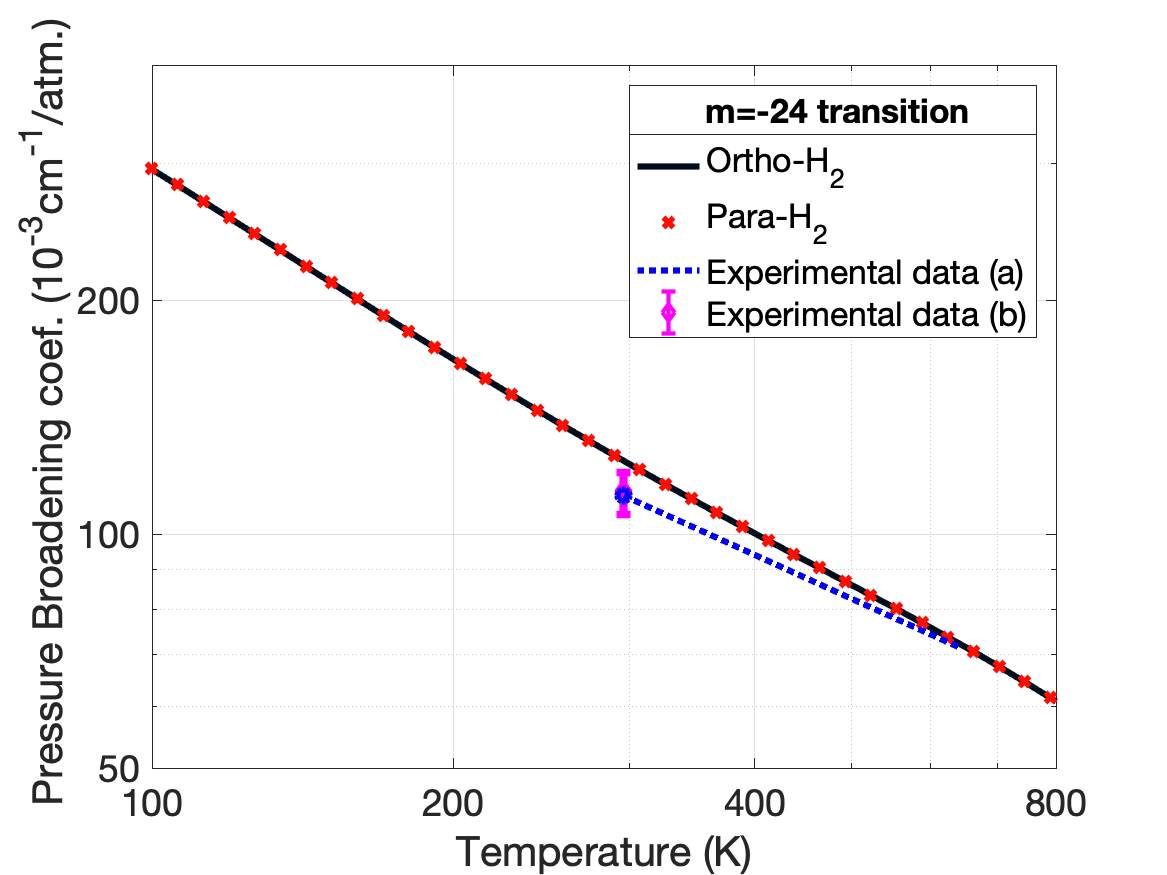}
    \caption{Theoretical pressure broadening coefficient $\gamma(T)$, for the $m=-24$ (P(24)) transition. The results for collisions with \textsl{ortho}-$\mathrm{H}_2,\,j_2=1$ is shown in black, for \textsl{para}-$\mathrm{H}_2,\,j_2=0$ in red. (a) Experimental point and temperature dependence from \cite{hanson2014}; these experiments were performed from 300K to 640K. (b) Experimental point from \citet{padmanabhan2014}.}
    \label{fig:results}
\end{figure}

The cross-sections $\sigma(E_{coll.}; m=-24)$ (elastic, inelastic, and pressure broadening, see appendix) were computed for about 100 collision energies, ranging from 10 to 1,500 $\mathrm{cm^{-1}}$. We present our finding in \autoref{fig:results}, which shows that the precision requirements are achieved. For 296~K, the theory ($\gamma_{\text{theo}}= 0.120\mathrm{cm^{-1}/atm.}$) slightly overshoots  the experimental values, $\gamma_{\text{exp.}}=0.112 \text{ or } 0.113\mathrm{cm^{-1}}$/atm). The difference ($6.9\%$) is better than the precision requirement defined in \autoref{sec:requirements}. In \citet{hanson2014},  the temperature-dependent slope lacked error bars, and therefore should only be interpreted as an approximation.

\section{Discussion \& Conclusion}
\label{sec:conclusion}

\subsection{Caveats and Previous State-of-the-art}
Computations of the pressure broadening coefficients for the aforementioned transition (i.e.  $m$ = -24) are performed through a series of codes that enables a full quantum dynamical approach to the problem, with \emph{no adjustable free parameter} (details in \autoref{sec:quantum}). While the underlying theoretical framework was presented decades ago  \citep[e.g.,][]{1958PhRv..112..855B,1981AdPhy..30..367P,GREEN1977119}, it is still able to provide accurate calculations today (\cite{Hutson2019}; Jaïdane et al., in prep.). Indeed, while we assessed that a 10\% precision of the line broadening parameters is required for state-of-the-art applications in exoplanetary sciences, we found that, for the transition computed, the agreement with existing measurements is well within this requirement ($\sim$7\%). While our calculations do not reach the accuracy achieved by the Torun group, who also compute the non-Voigt parameters \citep[e.g.,][]{2021JQSRT.27207807S, Gancewski2021}, our calculations are much faster (i.e., scalable) and can maintain a degree of accuracy required by atmospheric modeling applications.

Yet, a few remarks are in order. The calculated outputs for this particular transition (i.e. $m$ = -24 transition) exhibit only a weak dependence on any input parameter (except for the convergence of the quantum base, which remains paramount). For example, the $\sigma_{PB}(E)$ function displays practically no structure nor dependence of the para or ortho character of \ce{H2}. This is plausible due to particular aspects of the calculation. The occurrence of large quantum numbers included in the calculation significantly increases the number of channels (i.e. the quantum state during or after the collision), reduces the quantum interferences, and enables a semi-classical behavior of all the angular algebra (see \citet{Schaefer:1987aa} and section \ref{sec:pbcalc} in the appendix).

The complete \textsl{ab initio} computation of pressure broadening coefficients has been undertaken, mostly with atomic colliders \citep{Thibault:2000aa,2022JChPh.156j4303G,2021JQSRT.27207807S} but also molecules \citep{Thibault:2011bb,Wiesenfeld:2013aa}. All these \textit{ab initio} calculations only require physical inputs for the experimental geometry of the target and projectile molecules and their experimental rotational constants, possibly from ground and vibrational excited states.  The actual dynamics of vibration (i.e., the full wave-function of a ro-vibrator) has not been included in the dynamics as the vibration plays more of a spectator role, as underlined by the very small cross-section that allows for vibrations to be excited or quenched \citep{Wiesenfeld:2022aa}.

\subsection{Scalability}

A fundamental advantage of the solution to the exoplanet opacity challenge presented here relates to its scalability. \textsc{YUMI}, our FORTRAN-based quantum scattering code, takes advantage of the CPU-architectures, in particular the extensive parallelism and memory hierarchies. The code has been optimized using an \textsc{OpenMP} framework, and streamlined for simultaneous multiple jobs submission leveraging toolkits such as \texttt{LLTriple}  \citep{byun2015} developed for handling thousands of jobs in parallel. Our code performed at $\sim$30 GigaFlops during the speed benchmark test performed in a MIT SuperCloud node, which consists of Intel(R) Xeon(R) Platinum 8260 CPU$@$2.40GHz with 48 cores. The current speed bottleneck is the matrix inversion required for solving Schr\"odinger equation (Jaïdane et al. in prep., see also  Sec. \ref{sec:quantum}), which involves a recursive algorithm implemented from \citet{matrixinversion}, built on top of \texttt{Intel's} \textsc{MKL} library. Our algorithm scales at $\mathbf{O}(N ^{\sim 2.6})$, i.e. smaller than 3, with $N$, the matrix  size. For \ce{CO2-H2}, the typical matrix size is $\sim$1000, and the typical calculation takes less than $\sim$1 hour for a single transition at a particular energy. For a single transition, such calculations needs to be performed roughly at 100 different energies, and for more complex molecules such as water, matrix sizes reach $\sim$10,000. The related calculations would thus take $\sim$400 times more computational time (i.e. $\sim$2 weeks), significantly longer but still within our computational reach. Leveraging the symmetry within the molecule, GPUs, and possibly machine-learning techniques in future applications will further help us enhance the scalability of our approach.

\subsection{Future Prospects}

The results presented in this study provide a proof-of-concept for a scalable solution to a key limitation behind the exoplanet opacity challenge, namely the modeling of broadening. While our application focuses on the data-scarce CO$_{2}$-H$_{2}$ collisional system due to its importance in most of the exoplanet spectra acquired with JWST and the scarcity of data available for the system, it can now be applied to the hundreds of other collisional systems of relevance for JWST and future missions.

While the CO$_{2}$-H$_{2}$ system could be investigated with sufficient funding and careful experimental works on an adequate timescale ($\sim$ 1\,yr), it is just one collisional system out of thousands of molecules of relevance. Unfortunately, the measurements of broadening parameters with sufficient accuracy are challenging, and often rendered impractically complex and expensive due to safety issues (as a reminder, the primary targets of exoplanetary sciences remain hydrogen-rich high-temperature worlds). Therefore, the methodology presented here provides a timely and realistic approach to fulfilling the opacity needs of the exoplanet community, and helps pave the way toward robust atmospheric inferences reaching accuracy driven by the instrument itself (rather than opacity models).

\software{\textsc{emcee} \citep{emcee},  \textsc{MOLPRO} \citep{MOLPRO}, \texttt{Matlab}, \textsc{tierra} \citep{Niraula2022}, \texttt{\textsc{YUMI} (Jaïdane et al. in prep.)} }

\section*{Acknowledgements}
JdW and LW thank the MIT Global Seed Funds and MIT-France for their support towards this study. NJ thanks Universit\'e Paris-Saclay for a visiting fellowship. IG and RH acknowledge NASA PDAR grant 80NSSC24K0080. LW and NJ thank Olivier Dulieu for fruitful discussions. LW, PN, and JdW thank Jeremy Kepner, Deborah Woods, and Cooper Loughlin for their insights regarding optimal deployment on supercomputers, scalability, etc. which will be pivotal in the next steps of this project.
Part of these computations were performed on the Jean-Zay IDRIS-CNRS supercomputer, contract A0140810769. Other parts were computed on the MIT-Supercloud/Lincoln-MIT computers. The authors acknowledge IDRIS and the MIT SuperCloud and Lincoln Laboratory Supercomputing Center for providing resources (HPC, software expertise, database, consultation) that have contributed to the research results reported here.

\clearpage

\appendix

\restartappendixnumbering
\section{Notations}\label{sec:notations}

We used in this work the following conventions and notations:
\begin{description}

    \item [Quantum numbers and spectroscopy] To avoid any ambiguity between commonly used notation for angular momenta, we use the following notation: $j_1$, $j_2$, angular momentum resp. of the \ce{CO2} or \ce{H2} molecule;  $l$, orbital angular momentum of the collision; $J$, total angular momentum of the collision (a conserved quantity for a collision). The running index $m$ is defined as follows $m =-j_1''$ for P branch, $m=+j_1''$ for he Q branch, and $m=+j_1''+1$ for the R branch. Therefore for the transition studied here $m$=-24 corresponds to P(24). Primed and double primed quantities are lower and and upper levels, respectively. If one looks at the HITRAN file for carbon dioxide they will see branch designation and the $j_1$'' value for unique identification. \newline
    Vibrational levels of CO$_2$ are given (in HITRAN notation) by $[v_1 v_2 l_2 v_3 q]$, where $v_i$ are the three normal modes, i.e. symmetric stretch, bend, and asymmetric stretch, $l_2$ is the rotational vibration quantum number and $q$ labels the levels of the same polyad.
    \item [Energy] Energies are defined as follows: (i) $E$ is the total energy of the collision, (ii) collision energy $E_{coll.} = E -E_{\text{rotation}}(\ce{CO2}) - E_{\text{rotation}}(\ce{H2})$.
\end{description}

\section{Quantum dynamics \hfill}\label{sec:quantum}
Our aim is to compute pressure broadening coefficients for the 4.3~$\mu$ transition of $^{12}$C$^{16}$O$_{2}$, that is the IR lines that correspond to
$v''_3=1, \, j_1'' =j_1' \pm 1\leftarrow v'_3=0,\, j_1'$, broadened by collisions with \ce{H2}. We aim at a full quantum \textsl{ab initio} scheme, based on the close coupling (CC) formalism \citep{1960RSPSA.256..540A,1952RvMP...24..258B,1958PhRv..112..855B}. In order to do so, one needs successively: (i) a series of \textsl{ab initio} potential energy points $V(\mathbf{R}_n) $ describing the interaction of \ce{CO2} with \ce{H2} ($\mathbf{R}$, the 4 coordinates setting the geometry of \ce{CO2} with respect to \ce{H2}); (ii) a fit of these $V(\mathbf{R}_n) $ points onto a suitable functional form valid ; (iii) solving the quantum dynamical equation of the collision, in order to get the \textsf{S}-matrix connecting the entrance and outgoing channels of the collision (usually, solving the Schr\"odinger equation); (iv) from the knowledge of the \textsf{S}-matrix, constructing the pressure broadening cross-sections and rate coefficients. This section describes the details of all four steps.

\subsection{Potential Energy \textsl{ab initio} points}\label{sec:pes}

\begin{figure}[ht!]
   \centering
   \includegraphics[width=0.5\linewidth]{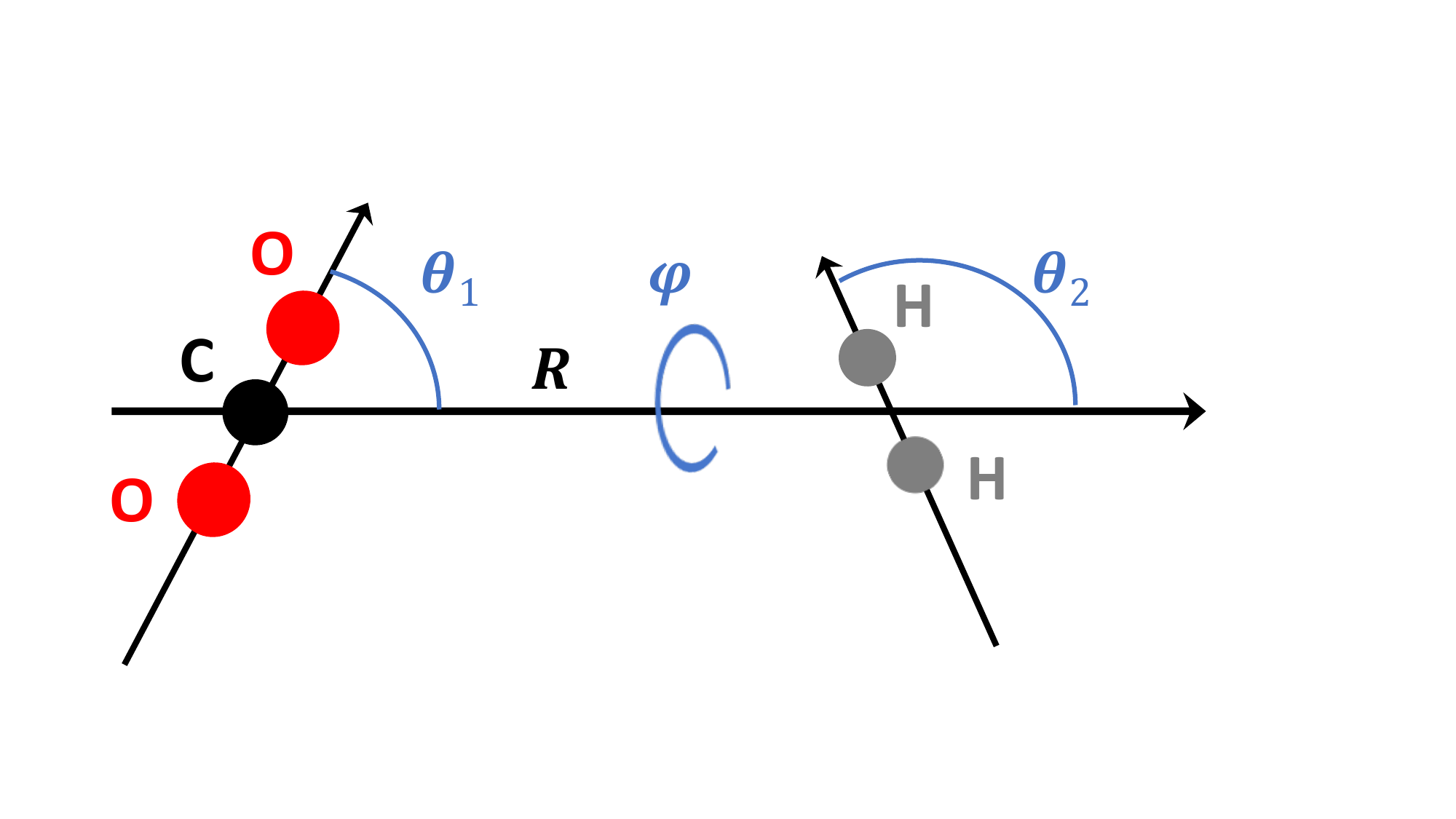}
\caption{The 4 degrees of freedom ($R$, $\theta_1$, $\theta_2$, $\phi$) of the \ce{CO2}--\ce{H2} complex, in the rigid bodies approximation. }
    \label{fig:geometry}
\end{figure}

We compute the pressure broadening by \ce{H2} of  Infra-Red transitions of \ce{CO2}. Hence, it would seem sensible to allow for \ce{CO2} vibrational modes to be included in the Potential Energy Surface (PES). However, vibrational motion influences very weakly the pressure broadening coefficients \citep{hitran2020,2022ApJS..262...40T}, at least in the temperature regimes that we aim at, \(50 \leq T\leq 1000 \,\mathrm{K}\). Note
that the same does  \emph{not}  seem to be true for line shifts. We are thus entitled to restrain our computations to rigid body interaction. We use the ground state vibrational averaged geometries for \ce{CO2} and \ce{H2}, with $r(\ce{CO})= 2.1944$~bohr \citep{2022JChPh.156j4303G} and $r(\ce{HH})= 1.44836 $~bohr \citep{Sahnoun2020}. The
PES has 4 degrees of freedom, namely, the $R$ distance between the center of masses of the two molecules, the $\theta_1$ and $\theta_2$ angles that describe the orientation of the two linear molecules with the intermolecular axis and $\phi$, the dihedral angle, see figure \ref{fig:geometry}.

In the vibrational ground state, both molecules display  $D_{\infty h}$ symmetry; the ground state intermolecular  PES must reflect these symmetries. With the geometry defined in figure \ref{fig:geometry}, the whole PES is described by  $0 \leq
\theta_1 < \pi$, $0 \leq \theta_2 < \pi/2$, and $0\leq \phi < \pi/2$.

As the atoms are relatively light, and as we have to deal with  22 electrons only (among those, 6 core electrons), it is possible to compute the PES in multiple ways, of various precision. This allows us to study the influence of the PES precision on the observables, elastic and inelastic cross-sections, and pressure broadening rate coefficients in particular.

\subsubsection{\textsl{Ab initio} quantum chemical computations}\label{sec:method_ab_initio}
The interaction energy is computed via the super-molecule approach, including systematically the basis set superposition errors (BSSE, \citep{2019JChPh.151g0901K}, see also our previous works \citep{Khalifa:2019ac, Sahnoun:2018aa} for a detailed description). All computations are performed at the CCSD(T) level, with the 2020 version of the \textsc{MOLPRO} code \cite{MOLPRO}.
The interaction energy is defined as:
\begin{equation}\label{eq:bsse}
    E^{AB}_{\text{interaction}} = E^{AB}_{\text{Dimer }AB} - E^{AB}_{\text{Monomer }A} - E^{AB}_{\text{Monomer }B}\quad ,
\end{equation}
were the superscript $AB$ denotes the full molecular basis $A \oplus B$, $A\equiv \ce{CO2}$ and $B\equiv \ce{H2}$.

In order to evaluate the precision needed for computing elastic and inelastic cross-section for the collision \ce{CO2 - H2}, we used four precision levels for computing the interaction (all described in the documentation of Molpro, \citep{MOLPRO}: (1) basis set aug-cc-pVTZ (denoted in the following
by \avtz), alongside with a bonding function center \citep{Shaw:2018aa} at 2/3 of the distance between the center of mass of \ce{H2} and the nearest atom of the \ce{CO2} molecule (The PES  depends very weakly to the actual position of the bonding function, as long as it is not too nearby any actual atom). (2) The same formalism but with a basis set aug-cc-pVQZ (denoted in the following by \avqz). (3) An extrapolation to the complete basis set, with the \avtz~and \avqz~values at each \textsl{ab initio} point, and an extrapolation formula \citep{Varandas:2021aa}:
\begin{equation}
V_{\text{cbs}} = \frac{3^3\cdot V_{\text{avtz}}-4^3\cdot V_{\text{avqz}}}{3^3-4^3}
\label{eq:extrap}
\end{equation}
The \textsl{ab initio} PES obtained via formula \eqref{eq:extrap} is called hereafter \avcbs.

The last formalism (4) is the so-called `gold standard', widely used for computing intermolecular interaction, the CCSD(T)-F12a formalism \citep{2019JChPh.151g0901K}. We used exclusively the \avqz \ basis for this computation (even if \avtz~could yield very similar results). Using a bonding function proved unreliable and was not used. We noted the well-known problem of size inconsistency of the explicitly correlated F12 family of methods. We duly subtracted the value of the interaction at infinity (here, 50~bohr), which was isotropic within less than 1\%.

Some points (see below) were computed at the aug-cc-pV5Z level (\avcz), in order to have some hints of the convergence of the PES, on the one hand, and to see if the CCSD(T)-F12 methods are reliable, especially so for overshooting the minimum of the PES. Results around the different stationary points are shown in table \ref{tab:pot_min}. As briefly explained beforehand, convergence at the wave-number scale is essentially achieved for \avcz~and the more economical \avcbs. \avqz~method with the F12 acceleration slightly overshoots the values, and \avtz~method is clearly less precise. However, all methods agree one with another within 4-5\% and with the earlier computation of \citet{Li:2010aa}.

\subsection{Geometries and fit}\label{sec:geom_fit}
While some formalism allows for computing the PES with a limited number of points \citep{2020PCCP...2217494B,Ajili:2022aa}, we decided to compute the PES on a dense grid of \textsl{ab initio} points, like we did repeatedly in previous cases, like \cite{Sahnoun2020,Valiron:2008aa,Rist:2012aa,Khalifa:2020aa}. To summarize, we compute \textsl{ab-initio} points for 500 sets angles \(\left\{\theta_1,\,\theta_2,\,\phi\right\}\) per distance $R$.
These angles are randomly chosen with $0\leq \theta_{1,2} < \pi$ and $0\leq \phi< 2\pi$. In doing so, we do not impose the $C_{\infty h}$ symmetries of \ce{H2} and \ce{CO2}. The subsequent fit of the potential should reveal the symmetries (see below). We computed the potential for 29 distances, between 4.20~bohr and 50.00~bohr. Note that the random set of angles is the same for each distance, facilitating the fit.

We fit the PES distance by distance, with the following formula adapted from \cite{GREEN:1975ae}.
\begin{equation}
\label{eq:expansion}
V(R,\theta_1,\theta_2,\phi)=\sum_{l_1,l_2,L} V_{l_1,l_2,L}(R) X_{l_1,l_2,L}\left(\theta_1,\theta_2,\phi\right)
\end{equation}
with the angular functions $X$ equal to :
\begin{align}
\begin{split}
     X_{l_1,l_2,L}\left(\theta_1,\theta_2,\phi\right) = \frac{2}{\left(4\pi\right)^{3/2}}\,(-1)^{(l_1-l_2)}
   \left\{\begin{pmatrix}l_1 &l_2& L\\ 0 & 0 & 0 \end{pmatrix}\mathcal{P}_{l_1 0}(\theta_1)\mathcal{P}_{l_2 0}(\theta_2) \right.\\
 + \sum_{m=1}^{\min(l_1,l_2)}2(-1)^m
\left.\left [ \begin{pmatrix}l_1 &l_2& L\\ m & -m & 0 \end{pmatrix}\mathcal{P}_{l_1 m}(\theta_1)\mathcal{P}_{l_2 m}(\theta_2)\cos(m\phi)
\right ]  \right\}
\end{split}
\end{align}
with $(l_1+l_2+L)$ even and $\begin{pmatrix}
     . . . \\ . . . \end{pmatrix}$ are Wigner 3-$j$ symbols.The normalization is:
\begin{equation}
\int_0^\pi \mathcal{P}_{lm}(\theta)\mathcal{P}_{l'm}(\theta)\,\sin\theta\,\mathrm{d}\theta=
\delta_{ll'}
\end{equation}
We took great care to have the same expansions and normalizations of the angular functions $ \mathcal{P}_{lm}(\theta) $  in this expansion and in the scattering code. To avoid any confusion, the explicit relations between the $\mathcal{P}_{lm}(\theta)$
functions, the associated Legendre functions, and the spherical harmonics are given in \cite{Zare:1988aa}, section~1.3.

\begin{table}

    \centering
    \begin{tabular}{l|ccccc|c}
      Method           & \avtz   & \avqz &    \avcz & \avcbs & \textsf{avqz-F12} &\citet{Li:2010aa}\\
      & & & & & &\\
      \hline
      & & & & & &\\
     $R_{\text{min}}$(bohr)        &     5.63    &    5.605  &       5.5875   &     5.595 &  5.585    &5.612\\
    & & & & & & \\
        $V{_\text{min}}$(cm$^{-1}$)  & -215.5  &    -219.59   &      -221.97   &   -222.65 & -224.77  & -219.65  \\
  & & & & & & \\
    \end{tabular}
    \caption{Minimum $V_{\text{min}}$ of the \ce{CO2-H2} potential energy surface, at intermolecular distance
$R_{\text{min}}$. The geometry is always $\theta_1=90^\circ$, $\theta_2=90^\circ$ and $\phi = 0^\circ$. The partial \avcz \ computations are included. Not that our computations are at average ground state geometries and \citet{Li:2010aa} are at equilibrium geometries.}\label{tab:pot_min}
\end{table}

We tried three bases for the fit, indexed by the the maximum allowed values of the $l_1, l_2$ and $L$ indices in equation (\ref{eq:expansion}), the so called :
\begin{enumerate}[(i)]
    \item \textit{precise} expansion, with $l_1\leq 24$, $l_2\leq 6$, and $L\leq 26$, resulting in 158 terms in the expansion (\ref{eq:expansion}). Because of limited relevance, not all terms allowed by angular algebra are included.
    \item \textit{intermediate} expansion, with $l_1\leq 20$, $l_2\leq 4$, and $L\leq 22$, resulting in  76 terms.
    \item \textit{imprecise} expansion with $l_1\leq 20$, $l_2\leq 2$, and $L\leq 22$, resulting in  38 terms.
\end{enumerate}

\noindent In all cases, only even values of angular momenta are allowed, because of the symmetries of the molecules.

\subsubsection{Potential surface}
\begin{figure}
    \includegraphics[width=0.75\linewidth]{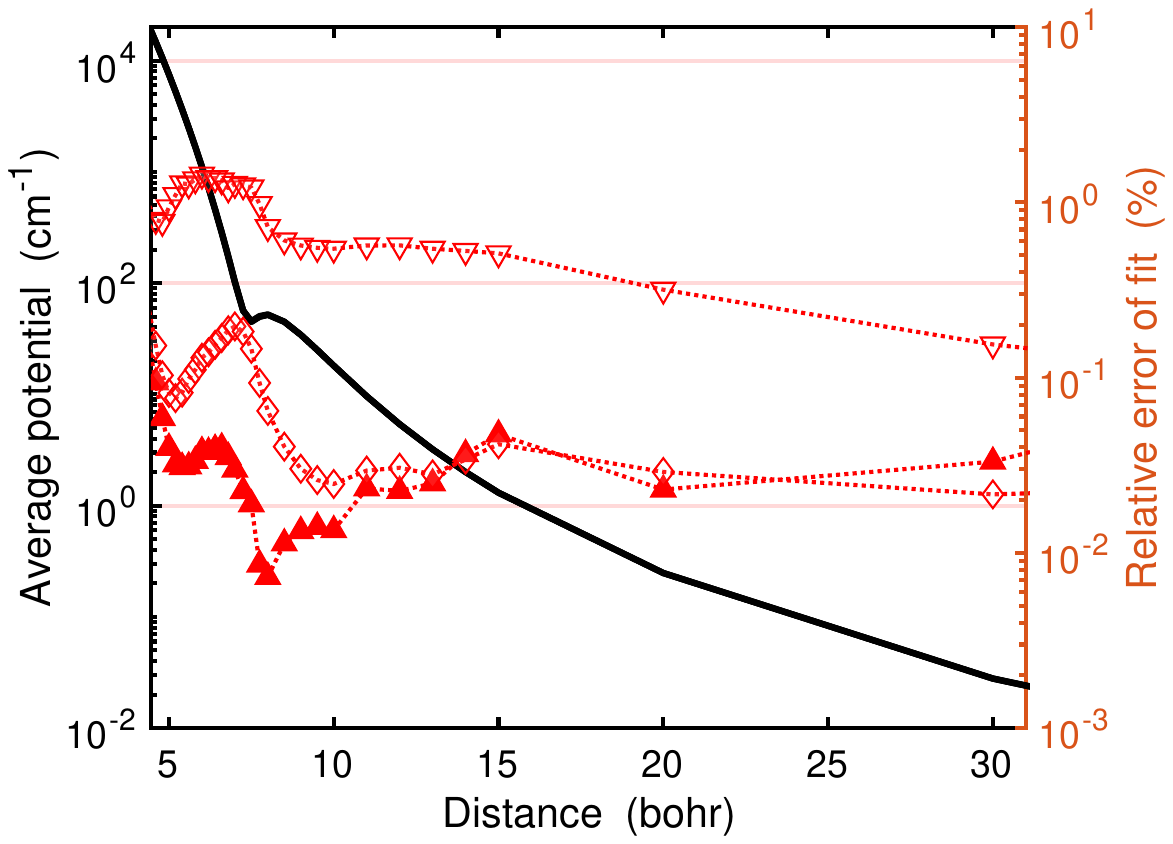}
    \caption{Different fits of the \textsl{ab initio} potential. The angle averaged absolute value potential, thick black line. Relative fit errors in red, scale on the right. Precise fit, filled upper triangles; intermediate fit, open diamonds; imprecise fit, empty lower triangles.}
    \label{fig:fits}
\end{figure}

We computed all 12 combinations \(\{\text{basis set}\}\times\{\text{fit precision}\}\), even if some combinations make little sense. For the \textit{precise} fit, comparison of some stationary points of the $V(R,\theta_1,\theta_2,\phi)$ potential functions are given in  \autoref{tab:pot_min}. \autoref{fig:fits} depicts errors in the fit for the three levels.

\subsection{Dynamics}
In order to solve  for the \textsf{S}- matrix characterizing the scattering, we resorted to the well-documented time-independent close-coupling (CC) formalism (see e.g. \cite{1960RSPSA.256..540A,Hutson2019,2023CoPhC.28908761A}). In order to accelerate computations and take advantage of the large memory available for each computing core, as well as of the massive now possible on modern supercomputers, we used a new home-written CC code. The main differences between the widely available  MOLSCAT code and the present one are the native use of OpenMP capabilities - readily extensible to GPU-based OpenACC capabilities\footnote{\texttt{\href{https://www.openacc.org}{https://www.openacc.org}}}, as well as memory management capable to take full advantage of the hierarchy of RAM in the cores and the nodes. A full description of the code will be available in a future publication.
\subsubsection{Cross-sections}
\begin{figure}
    \centering
    \includegraphics[width=0.75\linewidth]{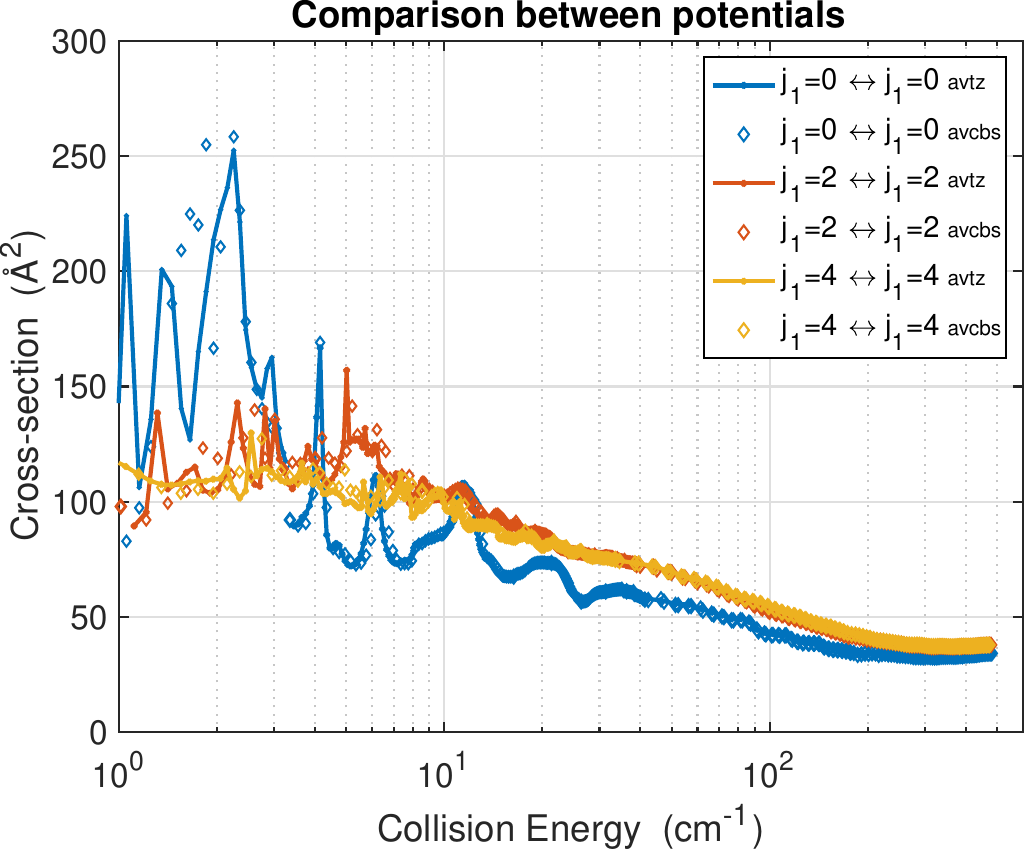}
    \caption{ Comparison between two \textsl{ab initio} bases, the less precise (\avtz \ basis), continuous line, and the most precise (\avcbs), diamonds. Large fit, with  156 terms. Elastic cross-sections for the collision \ce{CO2 -\textsl{ortho}\, H2}. $j_1$ is the \ce{CO2} rotational quantum number, and $v_3=0$.}
        \label{fig:comp}
\end{figure}

Once \textsf{S} or \textsf{T}  matrices are available, it is simple to sum up all relevant contributions and get the elastic/inelastic cross-sections at a given collision energy, $\sigma_{i j}(E_{coll})$, with $i \gtrless j$. These computations are much more sensitive to the details of the PES than their pressure broadening counterpart, as the next section will show.
An example of the sensitivity to  PES precision is given in figure \ref{fig:comp}.

\subsection{Pressure Broadening computation}\label{sec:pbcalc}
The knowledge of the \textsf{S} or \textsf{T}  matrices allows for the computing of the collision energy dependent pressure broadening $\sigma_{j_1''v''_b\leftarrow j_1' v'_b}(E_{PB})$. The formalism dates from \citet{1958PhRv..112..855B} and \citet{Schaefer:1987aa}.

\begin{figure}[h!]\label{fig:sec_pb}
    \centering
    \includegraphics[width=0.75\linewidth]{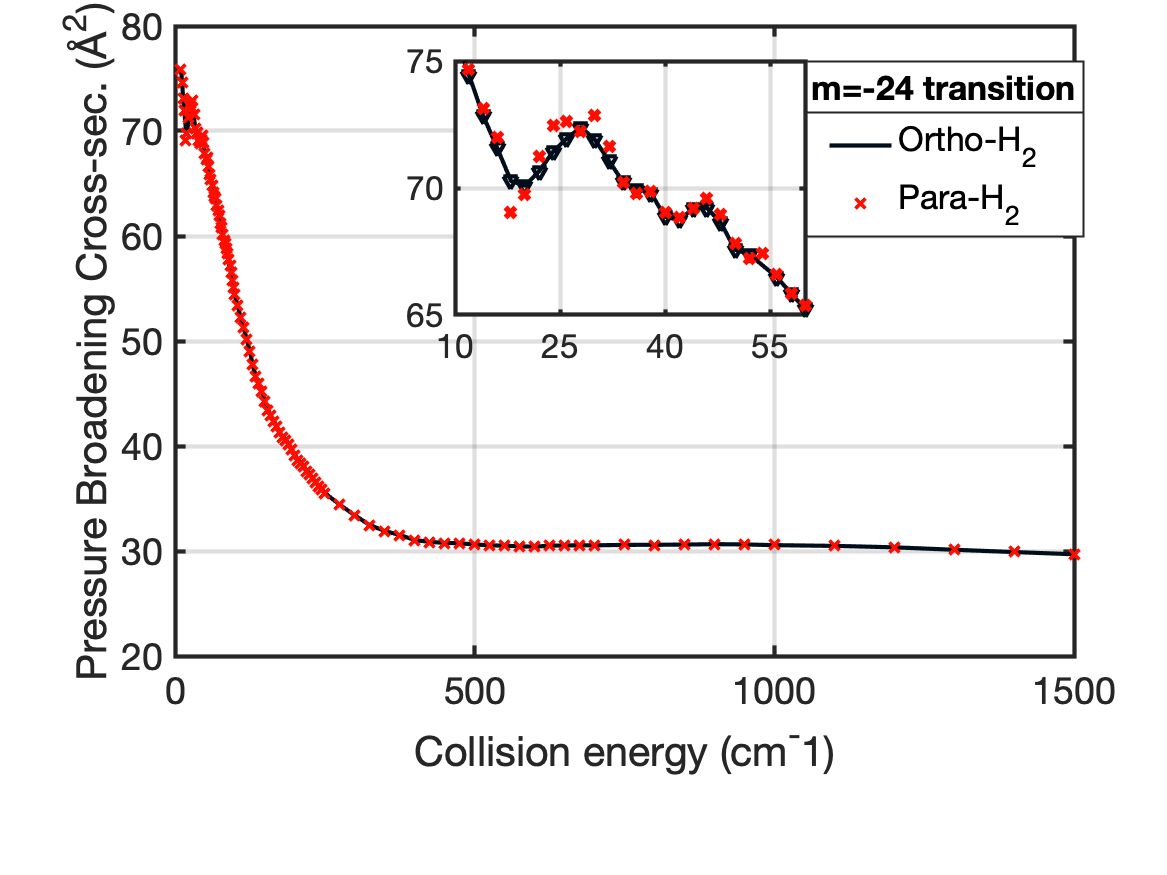}
    \caption{Pressure broadening cross-sections for the collision \ce{CO2-H2}, $\sigma{PB}(E_{coll})$. Ortho and para $\ce{H2}$ yield very similar results, except at very low energies (see inset).}
    \label{fig:pbsigma}
\end{figure}

The full analytical formula, eq.(1) of \citep{Schaefer:1987aa} was used in order to compute $\sigma_{PB}(E_{coll})$, for the transition $v''_3=0, j_1''=24 \leftarrow v'_3=1, j_1'=25$, at about 100 different collision energies $E_{coll}$. The degeneracy factor $(2j_2'+1)$ was taken into account, see remarks in \cite{GREEN:1975ae,Wiesenfeld:2013aa}. At each energy, 4 events were computed: collision with \ce{H2}, $j_2=0$, $j_1'=24 \text{ or } 25$, and the same with \ce{H2}, $j_2=1$. Results are presented  \autoref{fig:sec_pb}. The remarkable smoothness of the results contrasts with the multiple resonances observed in elastic cross-sections, like figure \ref{fig:comp}.

It must be underlined, though, that neither the smoothness of results nor the similarity of \textit{ortho}- and \textit{para}-$\ce{H2}$ pressure broadening cross-sections need to remain true for all regimes. Because of the presence of high angular momentum for \ce{CO2},
the collision studied here has a marked semi-classical character, witnessed by the behavior of $3-j$ and $6-j$ symbols. The behavior at low $j_1$ is expected to be much more quantum-like, as was observed repeatedly, for quadrupole dominated-collision e.g. \cite{Thibault:2011bb}, and even more so, for dipole-dominated collision \cite{Mengel:2000aa,2012PhRvA..86b2705D}.
From the knowledge of the $\sigma_{PB}(E_{coll})$, sections, we define the Maxwell-averaged cross-section and pressure broadening coefficient in the usual manner:
\begin{align}
   \sigma_{PB}(T)&=\frac{1}{\left(\kb T\right)^2}\,\int_0^\infty \sigma_{PB}(E_{coll})\;E_{coll}\,e^{(-E_{coll}/\kb T)}\, dE_{coll.} \label{eq:maxwell}\\
    \gamma(T)&=\frac{\bar v \sigma_{PB}(T)}{2\pi\,\kb T} \label{eq:gamma}
\end{align}
where the kinetic average relative speed is $\bar v= \sqrt{8\kb T/\pi\,M}$, $M$ being the reduced mass of the collisional system.

\subsubsection{Precision analysis}\label{sec:precision_analysis}

Given the acceptable error for retrievals (of the order of $\pm 10\; \%$), we found it useful to compare the results shown in figures~\ref{fig:sec_pb} and \ref{fig:results} with alternative methods, more economical in computer time.

The first test we made was to apply the Random-Phase Approximation (RPA) to the $\sigma_{PB}(E_{coll})$ computations, neglecting all interferences terms between the different channels \citep{1958PhRv..112..855B,Drouin:2012ab,2013JQSRT.116...79F}. It is known that this approximation (see e.g. the \ce{H2O-H2} case, \citep{2013JQSRT.116...79F} ) is valid only for collision energies $E_{coll}\gg \left|V_{min}\right|$, the minimum value of the potential. For the deep potential we have here ($V_{min} \simeq -220 \;\mathrm{cm^{-1}}$, see section \ref{tab:pot_min}), this approximation should not be not valid, and was indeed tested as being way too imprecise.

The second series of tests tried to disentangle the necessity of precise quantum mechanical methods for determining the potentials and precise fits of the potential points.
In order to shorten computation, we only computed $\sigma_{\mathrm PB}(E_{coll})$, for $E_{coll}=200 \,\mathrm{cm^{-1}} $  (about 300 K). Precision study is not expected to change significantly form a full-fledged $\gamma(T)$
computation, except for low temperature results, which are much more potentially dependent, because of the occurrence of scattering resonances \citep{2012PhRvA..86b2705D,Bergeat:2020aa}(see figure \ref{fig:comp}). Remarkably, the  $\sigma_{\mathrm PB}(E_{coll})$ value does not change at the 0.1\% level of precision for the different methods of computing the PES (\avtz\ up to \avcbs).

Changes in the fitting procedure precision do have a limited influence on the results, of at most 2\%. Fits with very few coefficients may sometimes make the dynamics diverge, and lead to nonphysical results; a compromise is necessary. However, it must be underlined that pressure shifts, inelastic and elastic cross-sections are much more sensitive to the details of the PES and fit. Also, there is now no indication if smaller values of $|m|$ behave the same way \citep{Thibault:2011bb}. Consequently, all data aside from pressure broadening should be carefully analyzed for convergence, and influence of the vibrational motion. Recall also that the inelastic cross-sections averaged over the Maxwellian distribution of collision energy yield the population transfer rate coefficients, \cite{1960RSPSA.256..540A,Roueff:2013aa}, useful for level population determination.

The relative insensitivity of $\sigma_{PB}(E_{coll})$ (hence of $\sigma_{PB}(T)$) to the details of the quantum computation comforts the idea of scalability of this computational scheme, thanks to the possibility of using properly tailored computations. It may even be envisioned, for lower quality results, to resort to approximate quantum scattering dynamical methods, based on Born first or higher order methods, and resorting to specific tests to validate approaches. For high enough $|m|$ values, the insensitivity to the rotational state of \ce{H2} halves the computing time. The question of classical simulation \citep{2018JQSRT.213..178H} of the dynamics is also to be carefully examined, from the point of view of precision and scalability. It must be recalled, though, that for lower values of $j_1'$ and $j_1''$, some of the present conclusions do not hold as we remain in the quantum-dominated regime, where the whole quantum machinery must be performed.

\bibliography{./Bibfiles/bibliography.bib}

\end{document}